\documentclass[iop,apj,tighten,twocolappendix,numberedappendix]{emulateapj}
\usepackage{apjfonts}

\usepackage{graphicx}
\usepackage{amsmath}
\usepackage{amssymb}
\usepackage{color}
\usepackage{mathtools}

\usepackage[breaklinks,colorlinks,citecolor=blue,linkcolor=red]{hyperref} 
\usepackage[all]{hypcap}

\newcommand\msun{\, \rm M_\odot}

\newcommand\kms{\, \rm km\,s^{-1}}

\newcommand\msmbh{{M_{\rm SMBH}}}

\newcommand\be{\begin{equation}}
\newcommand\ee{\end{equation}}

\begin{document}

\title{Gravitational-wave captures by intermediate-mass black holes in galactic nuclei}

\author{Giacomo Fragione\altaffilmark{1,2}, Abraham Loeb \altaffilmark{3}, Kyle Kremer\altaffilmark{1,2}, \& Frederic A.\ Rasio\altaffilmark{1,2}}
 \affil{$^1$Center for Interdisciplinary Exploration \& Research in Astrophysics (CIERA), Evanston, IL 60202, USA} 
  \affil{$^2$Department of Physics \& Astronomy, Northwestern University, Evanston, IL 60202, USA}
  \affil{$^3$Astronomy Department, Harvard University, 60 Garden St., Cambridge, MA 02138, USA}

\begin{abstract}
Intermediate-mass black holes (IMBHs) have not been detected beyond any reasonable doubt, despite their potential role as massive seeds for quasars and sources of tidal disruption events, ultra-luminous X-ray sources, dwarf galaxy feedback, and hypervelocity stars. Gravitational wave (GW) observations can help to find and confirm the existence of IMBHs. Current and upcoming detectors, such as LIGO, Virgo, KAGRA, LISA, ET, and DECIGO promise to identify the full range from stellar-mass to supermassive black holes (SMBHs). In this paper, we address the question of whether IMBHs can produce GWs in galactic nuclei. We consider the possibility that stellar black holes (SBHs) form bound systems and later coalesce with an IMBH through gravitational captures in the dense nucleus. We show that this mechanism is efficient for IMBH masses in the range $\sim 3\times 10^3\msun$--$2\times 10^4\msun$. We find that the typical distributions of peak frequencies and merger timescales depend mainly on the IMBH mass. In particular, the typical peak frequency is about $0.2\,$Hz, $0.1\,$Hz, $0.09\,$Hz, and $0.05\,$Hz for $M_{\rm IMBH}=5\times 10^3\msun$, $8\times 10^3\msun$, $1\times 10^4\msun$, and $2\times 10^4\msun$, respectively. Our results show that, at design sensitivity, both DECIGO and ET should be able to detect these IMBH--SBH mergers. Furthermore, most of the mergers will appear eccentric ($e \gtrsim 0.1$), providing an indication of their dynamical origin.
\end{abstract}


\section{Introduction}
\label{sect:intro}

The possible existence of intermediate-mass black holes (IMBHs) is one of the unsolved questions of modern astronomy \citep{mezcua2017}. IMBHs have masses in the range $\sim 10^2\ \mathrm{M}_{\odot}\lesssim M\lesssim 10^5\ \mathrm{M}_{\odot}$, higher than stellar black holes (SBH) and lower than supermassive black holes (SMBH). While the latter two families have direct proof for their existence \citep{korm2013,ligo2019}, there is only circumstantial observational evidence for IMBHs \citep{baldass2018,chili2018,lin2018}. Owing to their potential role in a wide range of phenomena, including the origin of SMBH seeds and galaxy evolution \citep{madau2001,tagawa2019}, tidal disruption events \citep{chen2011,fragle2018}, gravitational wave emission \citep{gair2011,fragl2018b}, ultra-luminous X-ray binaries \citep{kaaret2017ARA&A..55..303K}, such as HLX-1 in ESO243-49 \citep{farrell2009}, dwarf galaxy feedback \citep{silk2017}, and hypervelocity stars \citep{yut2003,levin2006,rassk2019}, finding observational imprints of the origin and evolution of IMBHs has recently attracted significant attention \citep*[see][for a review]{greene2019}. 

There are at least three main pathways to form IMBHs. The first mechanism involves the collapse of massive Pop III stars. Due to inefficient cooling, Pop III stars of a few hundreds solar masses collapse to an IMBH of $\sim 100 \msun$ \citep{madau2001,bromm2004,fryer2001,bromm2013,loebfur2013}. The second channel predicts that an IMBH of a very high mass ($\sim 10^4$--$10^6\msun$) may be born following the collapse of a gas clouds, without passing through all the phases of stellar evolution \citep{loeb1994,bromm2003,begelm2006}. IMBHs with masses in between these two extremes can be produced through gravitational runaway events in star clusters \citep{por02,gie15}. In this contest, repeated mergers of massive stars \citep{gurk2004,frei2006,panloeb2012} or stellar black holes \citep{mil02b,antoras2016} can give birth to an IMBH with mass $\sim 10^3$--$10^4\msun$. Other possibilities include the fragmentation of SMBH accretion disks \citep{McKernan+2012,McKernan+2014} and super-Eddington accretion onto SBHs in SMBH accretion disks \citep{koc11}. 

Recent efforts have been directed towards understanding all the possible observational imprints of IMBHs, which could be detected in a number of different ways. Accreting IMBHs could be found from radio to X-ray in galactic nuclei \citep{greene2007,baldass2018,chili2018} or as ultraluminous X-ray sources in the field \citep{kaaret2017ARA&A..55..303K}. The presence of dormant IMBHs can be inferred from stellar and gas dynamical searches both in galactic nuclei and globular clusters \citep{gual2010,girma2019,baum2019}. In these environments, IMBHs can also disrupt stars, resulting in detectable tidal disruption events \citep{fragle2018,fragleiginkoc18,lin2018}. The disruption of a white dwarf is of particular interest, since such an event is luminous only for IMBHs with masses $\lesssim 10^5\msun$, which have a Schwarzschild radius smaller than the white dwarf disruption radius \citep{rossw2008,rossw2009,shen2019,peng2019}.

Gravitational wave (GW) astronomy will help in the hunt for the first IMBHs to be discovered beyond any doubt \citep{kons2013,arca2019}. IMBH--SBH binaries may form in the cores of star clusters or in galactic nuclei, and may merge as intermediate mass ratio inspirals \citep[IMRIs;][]{seoane2007,mandel2008,fragleiginkoc18}. Present and upcoming GW observatories, including LIGO\footnote{\url{http://www.ligo.org}}, the Einstein Telescope\footnote{\url{http://www.et-gw.eu}} (ET), and LISA\footnote{\url{https://lisa.nasa.gov}}, will be able to detect GW sources from IMBHs of masses up to $\sim 100-1000\msun$, $\sim 10^3-10^4\msun$ and $\gtrsim 10^4\msun$, respectively \citep{bello2019}. Another third-generation mission, DECIGO\footnote{\url{http://tamago.mtk.nao.ac.jp/decigo/index_E.html}}, could reveal GW events across most of the IMBH mass spectrum. Using the non-detection of massive binaries in the first two observational runs, the LIGO/Virgo collaboration placed upper limits on merging IMBHs, of the order of $\sim 0.1$--$1$ Gpc$^{-3}$ yr$^{-1}$ \citep{ligov2019}.

Here we address the question of whether IMBHs can produce observable GW sources in galactic nuclei. Specifically, we consider the possibility that SBHs form a binary with an IMBH as a result of gravitational bremsstrahlung, with the binary later merging as an IMRI. While this process has been widely discussed in the context of SBH-SBH captures in galactic nuclei and star clusters \citep{oleary2009,gondan2018,rasskoc2019,sams2019}, the IMBH regime has received less attention. 

The paper is organized as follows. In Section~\ref{sect:formation}, we discuss how IMBHs form and migrate in galactic nuclei. In Section~\ref{sect:captures}, we discuss the process of capture through gravitational bremsstrahlung. In Section~\ref{sect:montecarlo}, we describe our Monte Carlo framework and we derive the typical GW signals as a function of GW peak frequency and GW strain in Section~\ref{sect:multiband}. We estimate the IMBH--SBH merger rates from this process in Section~\ref{sect:rates}. Finally, we discuss the implications of our findings and draw our conclusions in Section~\ref{sect:conc}.

\section{Intermediate-mass black holes in galactic nuclei}
\label{sect:formation}

Several mechanisms exist that could create IMBHs in galactic nuclei. They can either form ex-situ or in-situ. In the former case, IMBHs have to be delivered to the innermost galactic regions by some dissipation mechanism. Below, we describe some of these scenarios in more detail.

Star clusters are promising environments for forming an IMBH. This would be natural assuming that the observed relation between SMBH mass and the velocity dispersion of stars around it holds also for IMBHs \citep{merritt2013}. A number of studies showed that the most massive stars may segregate and merge in the core of the cluster, forming a massive growing object that can later collapse to an IMBH \citep{por02,gurk2004,frei2006,gie15}. In this case, the typical IMBH mass would be in the range $\sim 10^3-10^4\msun$. Star clusters born in the innermost galactic regions could to inspiral into the centres of galaxies by dynamical friction on timescales much shorter than a Hubble time \citep{tremaine1975,capuzz2008,gne14}. If these GCs host central IMBHs, this process could efficiently deliver these IMBHs to galactic centers \citep*{gurk2005,mast14,agu18,fragk18}. In this scenario, the IMBH forms ex-situ and is delivered to the galactic nucleus by dynamical friction acting on his parent cluster.

A second ex-situ formation scenario for IMBH is due to minor mergers of galaxies \citep*{volon03}. Following the mergers of galaxies, IMBHs can be delivered to the proximity of the major galaxy nucleus, owing to three different processes \citep{merr2005}. First, the IMBH inspirals independently towards the centre of the gravitational potential via dynamical friction. This is followed by a stage where the system loses energy and angular momentum as a result of stellar gravitational slingshots \citep{quin1996,sesa2006,rassk2019}. This process eventually drives the system to the subsequent stage of energy loss due to emission of GWs \citep{merr2005,doso2017}. In this scenario, the IMBH is formed in a larger environment than a star cluster, as within a dwarf galaxy \citep{silk2017,chili2018}, and its mass would typically be in the range $\sim 10^3-10^5\msun$.

Another mechanism that produces and delivers IMBHs close to the galactic nucleus involves Pop III stars. In this scenario, IMBHs form as the remnants of the very first massive stars \citep{vol2008,volo2009}, while dynamical friction would then deliver some of them close to the SMBH within a Hubble time \citep{madau2001}. In this channel, the typical IMBH mass would be $\sim 10^2-10^3\msun$, but significant accretion could later increase it significantly.

A different formation scenario is that of a collapsing gas cloud, which forms a massive IMBH without passing through the phases of stellar evolution \citep{loeb1994,bromm2003}. This channel produces IMBHs of $\sim 10^4-10^6\msun$, but could only work at high redshifts, where the pristine gas can efficiently suppress cooling and fragmentation.

IMBHs could also form efficiently in-situ in the gaseous disks of active galactic nuclei (AGN) \citep{McKernan+2012,McKernan+2014}. If migration traps are present in the gaseous disk surrounding an SMBH, differential gas torques exerted on the orbiting SBHs will cause them to migrate towards a migration trap \citep{secunda18}. Turbulence in the gaseous disk can knock orbiting SBHs out of resonance, but allowing them to drift close to the trap and experience a close interaction with the first SBH. The interactions are dissipative due to the gas, and it is possible that SBH-SBH binaries form and merge repeatedly, thus forming an IMBH \citep{mcker2019,yang2019}. The masses of the IMBHs are poorly constrained in this scenario, and IMBHs could continue to accrete gas, thus increasing their masses considerably \citep{McKernan+2014}.

If a gaseous disk is not present, IMBHs can form though repeated mergers of stellar black holes in dense systems \citep{antoras2016}. Here, the requirement is that the host nuclear cluster is dense and massive enough to retain the merger remnant following its recoil kick due to asymmetric emission of GWs \citep{lou2008,lou2012,hof2016}. If the merger products are retained, they can form dynamically new binaries with SBHs and merge again, thus leading to a significant mass growth \citep{anto2019}. The typical IMBH mass would be $\sim 10^2-10^4\msun$.

\section{Gravitational bremsstrahlung}
\label{sect:captures}

We start by deriving the cross-sections for IMBH-bremsstrahlung in galactic nuclei.

We first consider the cross section for an IMBH of mass $M_{\rm IMBH}$ to undergo an encounter with an SBH of mass $m_2=q M_{\rm IMBH}$ ($M_{\rm IMBH}>m_2$), within a pericenter distance $r_p \le \mathcal{R}_{cap}$. We define $\mathcal{R}_{cap}$ to be the maximum distance below which $M_{\rm IMBH}$ and $m_2$ remain bound. In the gravitational focusing limit, the encounter cross-section is simply \citep{quinshap87},
\be
\sigma_{\rm cap}=2\pi G \frac{M_{\rm IMBH}(1+q)\mathcal{R}_{\rm cap}}{v^2}\ ,
\label{eqn:crosect}
\ee
where $v\sim (G\msmbh/r)^{1/2}$ ($r$ is the distance from the SMBH) is the local dispersion velocity. For an interaction to result in a GW capture, the energy radiated at the first pericenter passage from GW emission \citep{Turner1977},
\be
\Delta E_{\rm GW}=\frac{85\pi}{12\sqrt{2}}\frac{G^{7/2}}{c^5}\frac{M_{\rm IMBH}^{9/2}}{\mathcal{R}_{\rm cap}^{7/2}}q^2(1+q)^{1/2}\ .
\ee
has to be equal to the relative kinetic energy $(1/2)\mu v^2$, where $\mu$ is the reduced mass of the binary,
\be
\mu=M_{\rm IMBH}\frac{q}{1+q}\ .
\ee
This fixes the maximum pericenter distance $\mathcal{R}_{\rm cap}$ to,
\begin{eqnarray}
\mathcal{R}_{\rm cap}&=&R_{\rm S,1}\left(\frac{85\pi}{96}\right)^{2/7}\left(\frac{c}{v}\right)^{4/7}q^{2/7}(1+q)^{3/7}=\nonumber\\
&=&R_{\rm S,1}\left(\frac{85\pi}{96}\right)^{2/7}\left(\frac{c}{v_h}\right)^{4/7}\left(\frac{r}{r_h}\right)^{2/7}q^{2/7}(1+q)^{3/7}\ ,
\label{eqn:rss}
\end{eqnarray}
where $R_{\rm S,1}=2GM_{\rm IMBH}/c^2$ is the Schwarzschild radius of the IMBH. The semi-major axis and eccentricity of the new-formed binary can be derived as \citep{oleary2009},
\be
a=\frac{G M_{\rm IMBH}^2 q}{2|E_F|}
\label{eqn:semcap}
\ee
and
\be
e=\left(1-\frac{2|E_F|v^2 b^2}{G^2 M_{\rm IMBH}^3 q (1+q)}\right)^{1/2}\ ,
\label{eqn:ecccap}
\ee
respectively. In the previous equations, $E_F=\frac{1}{2}\mu v^2 - \Delta E_{\rm GW}$ and $b$ is the impact parameter, related to the pericenter of the orbit through (assuming gravitational focusing),
\be
r_p \approx \frac{b^2v^2}{2GM_{\rm IMBH}(1+q)}\ .
\ee
Therefore, the maximum impact parameter for a bremsstrahlung capture is for $r_p=\mathcal{R}_{\rm cap}$,
\be
b_{\rm max}\approx \frac{[2G\mathcal{R}_{\rm cap}M_{\rm IMBH}(1+q)]^{1/2}}{v}\ .
\label{eqn:bmax}
\ee
Requiring a minimum pericenter to avoid head-on collisions sets the limit for the minimum impact parameter \citep{gondan2018},
\be
b_{\rm min}=\frac{4GM_{\rm IMBH}(1+q)}{cv}\ .
\label{eqn:bmin}
\ee
The peak frequency at formation is \citep{wen2003}
\be
f_{\rm GW}=\frac{\sqrt{G M_{\rm IMBH} (1+q)}}{\pi} \frac{(1+e)^{1.1954}}{[a(1-e^2)]^{1.5}}
\ee
After the binary is formed, it evolves due to GW radiation reaction, and merges on a timescale \citep{peters64}
\be
T_{\rm GW,IMBHB}=\frac{5}{256}\frac{a^4 c^5}{G^3 M_{\rm IMBH}^3 q(1+q)}(1-e^2)^{7/2}\ .
\label{eqn:tgwimbhb}
\ee

\begin{figure*} 
\centering
\includegraphics[scale=0.55]{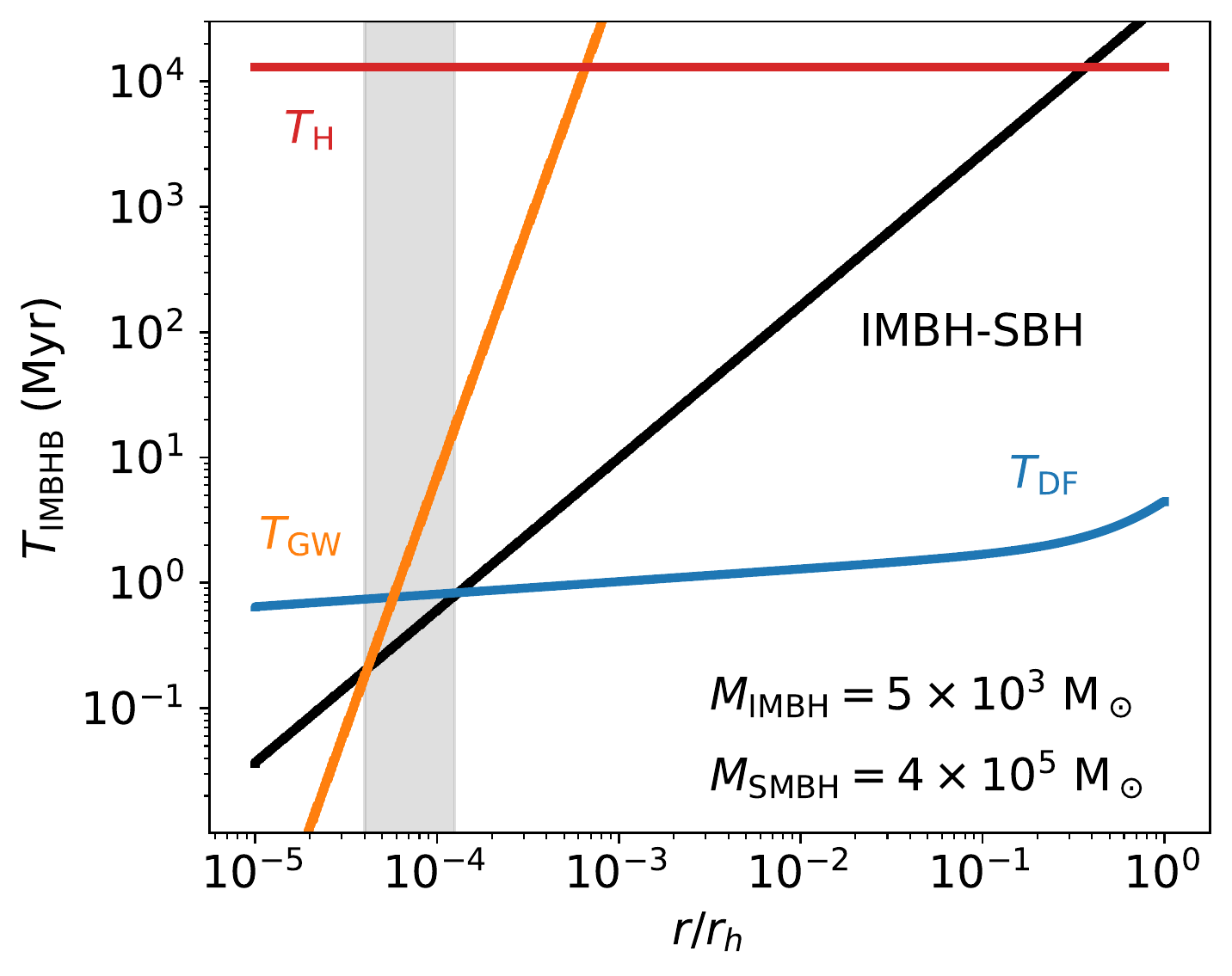}
\includegraphics[scale=0.55]{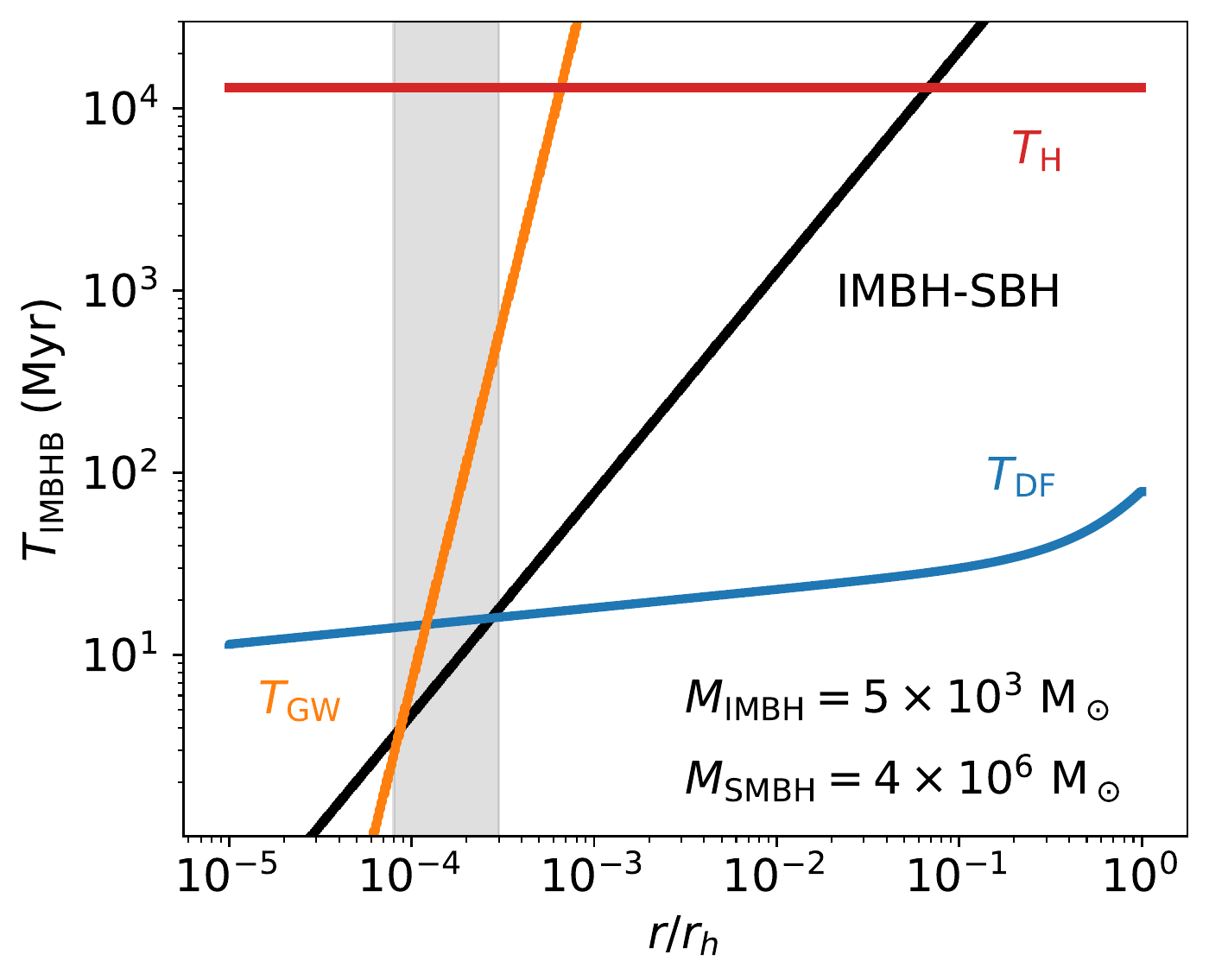}
\includegraphics[scale=0.55]{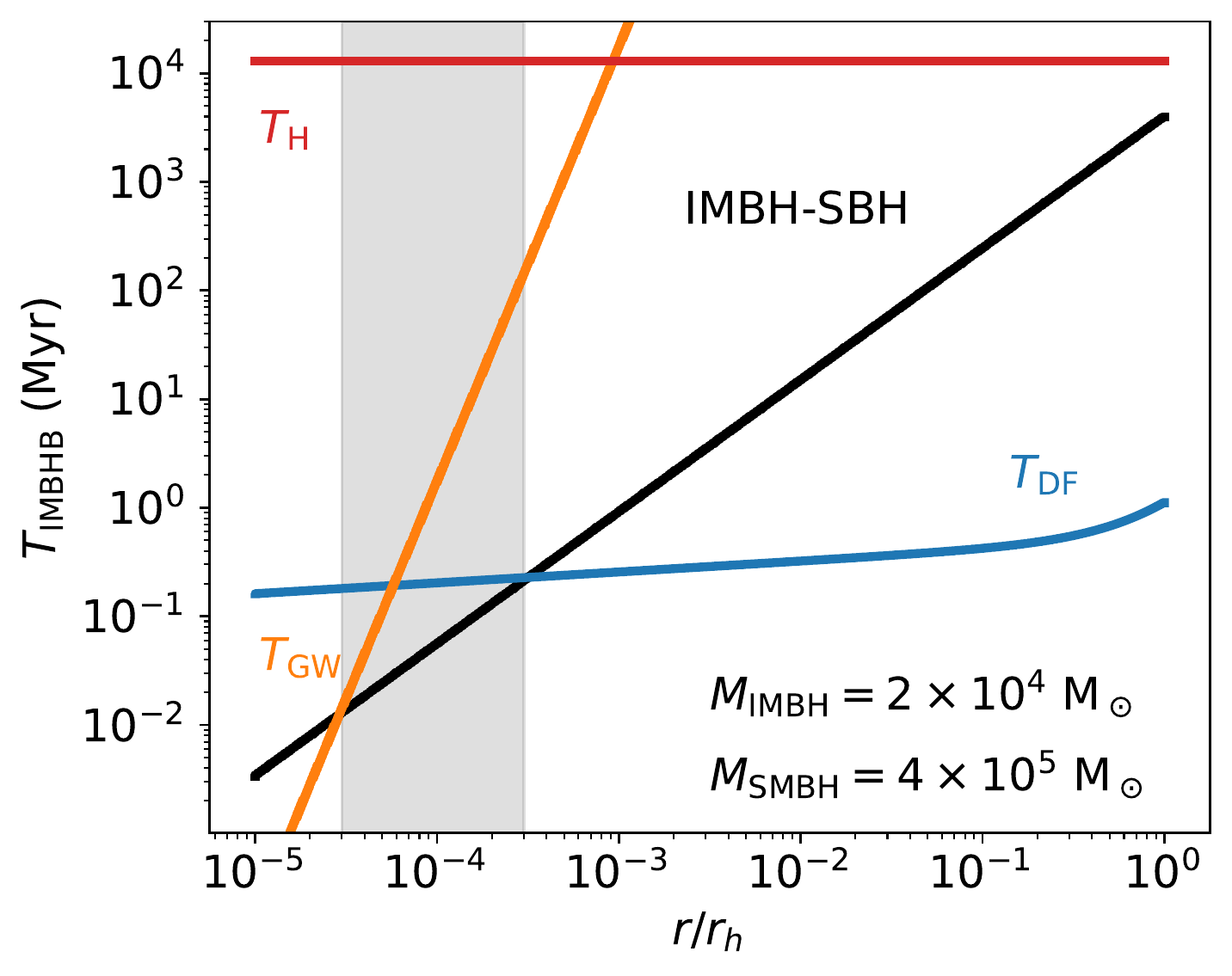}
\includegraphics[scale=0.55]{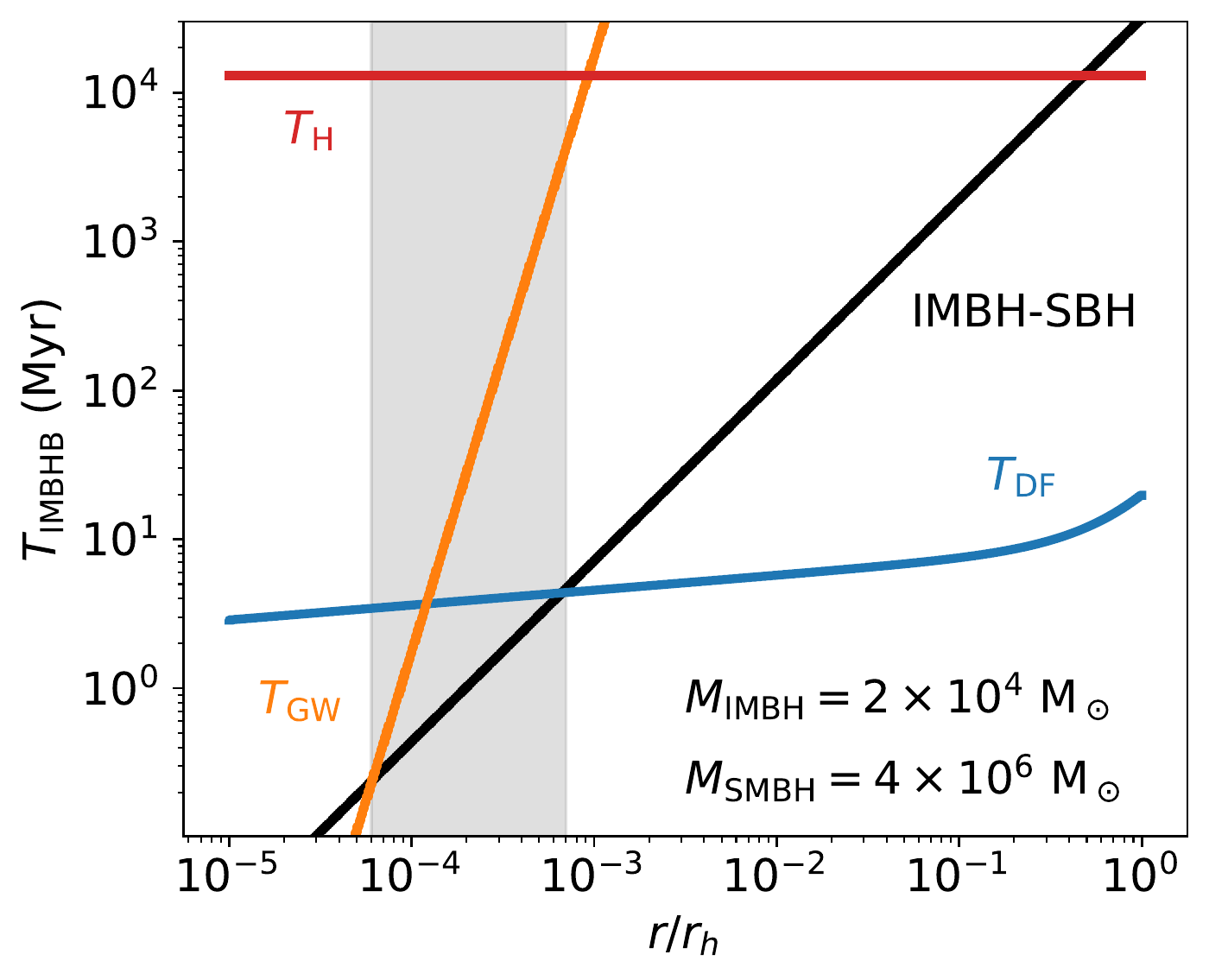}
\caption{Typical timescale (per IMBH) for the formation of IMBH--SBH binaries (Eq.~\ref{eqn:timbhb}) at a given position within the cusp, for different IMBH masses. Left: $\msmbh=4\times 10^5\msun$; right: $\msmbh=4\times 10^6\msun$. Shown in the figure also the dynamical friction timescale (blue line), the GW merger timescale for the SMBH-IMBH system (orange line), and Hubble time (red line). The shaded region represents the available portion of the phase space where IMBH--SBH binaries can efficiently form and eventually merge.}
\label{fig:timbhb}
\end{figure*}

To derive the relevant timescales, we need to quantify the stellar and compact-object populations within the SMBH sphere of influence. The sphere of influence is defined as the region within the characteristic radius \citep{merritt2013},
\be
r_h=\frac{G\msmbh}{v_h^2}\ ,
\ee
where $v_h$ is the galactic dispersion velocity at the radius of influence. A number of studies have found that stars and compact objects form cuspy profiles \citep{bahc1976,hopale2006,perets2007,aha2015,frasar2018}. For simplicity, we assume that the cusp is dominated by solar-mass stars, with density profile \citep{oleary2009,amarpreto2011},
\be
n_{\rm MS}=\frac{N_{\rm MS}}{A r_h^3}\left(\frac{r}{r_h}\right)^{-\alpha_{\rm MS}}\ ,
\ee
where $A=4\pi/3$, $\alpha_{\rm MS}\sim 1.6$, and $N_{\rm MS}$ is the number of objects at the influence radius given by \citep[see Eq.~9 in][]{gondan2018}, 
\be
n_{\rm MS}(r=r_h)=\frac{N_{\rm MS}}{A r_h^3}=1.4\times 10^5\ {\rm pc}^{-3} \left(\frac{10^6\msun}{\msmbh} \right)^{1/2}\ .   
\ee
The effect of different SMBH masses has been accounted for through the "M-sigma" relation \citep{tremaine2002},
\be
\msmbh=\eta v_h^4\ ,
\ee
where $\eta =4\times 10^{-2} \msun (\kms)^{-4}$ is a constant, and $v_h$ is the velocity dispersion at the influence radius $r_h$ of the SMBH. Using this "M-sigma" relation, $r_h$ and $T_h$ can be rewritten as,
\be
r_h=G\eta^{1/2} M_{\rm SMBH}^{1/2}\ ,
\ee
and
\be
T_h=\frac{r_h}{v_h}= G\eta^{3/4} M_{\rm SMBH}^{1/4}\ ,
\ee
respectively. We also consider an SBH population,
\be
n_{\rm SBH}=\frac{N_{\rm SBH}}{A r_h^3}\left(\frac{r}{r_h}\right)^{-\alpha_{\rm BH}}\ ,
\label{eqn:nsbh}
\ee
where $N_{\rm SBH}=0.023(3-\alpha_{\rm BH})/(3-\alpha_{\rm MS})N_{\rm MS}$ \citep{gondan2018} and $\alpha_{\rm SBH}\sim 2$ \citep{bw77,hopale2006}\footnote{For a mass-spectrum, the characteristic slope of the cusp of a given population depends on the mass of the object, the more massive the steeper \citep{kesh2009,alex2017}}.

Using Eqs.~\ref{eqn:crosect},\ref{eqn:rss},\ref{eqn:nsbh}, we can estimate the typical timescale (per IMBH) for the formation of an IMBH--SBH binary at a given position within the cusp,
\be
T_{\rm IMBHB}=\frac{1}{n_{\rm SBH} \sigma_{\rm cap} v}\ ,
\ee
where $n_{\rm SBH}$ is the number density of SBHs. Plugging the relevant parameters derived above, the typical timescale for the formation of a binary containing an IMBH can be rewritten as,
\be
T_{\rm IMBHB}=\frac{QA}{2\pi N_{\rm SBH}}T_h \frac{r_h}{\mathcal{R_{\rm cap}}}\left(\frac{r}{r_h}\right)^{\alpha_{\rm SBH}-1/2}\ ,
\label{eqn:timbhb}
\ee
where $T_h=r_h/v_h$ is the orbital period at the influence radius and $Q=\msmbh/(M_{\rm IMBH}+m_2)$.

An IMBH can capture SBHs whenever its inspiral time due to dynamical friction is long enough \citep{binntrem87}. We estimate the dynamical friction timescale following \citet{gurk2005},
\begin{equation}
T_{\rm DF}=\frac{[1+(4-\alpha)/(6-\alpha)\zeta](1+\zeta)^{1/2} Q T_h}{(3-\alpha) B \zeta}\left(\frac{r}{r_h}\right)^{3/2}\ .
\label{eqn:tdf}
\end{equation}
In the previous equation, $B\sim 1.7$ is a numerical factor of the order of unity that accounts for the dynamical friction coefficient and the Coulomb logarithm \citep{mcmil2003}, $\alpha\sim 1.6$ is the overall cusp slope\footnote{\citet{alexhop2009} have shown that, in the case of strong mass segregation, the dominant cusp of $1\msun$ stars would be distributed in a cusp with slope $\sim 1.4$, while white dwarfs, neutron stars and SBHs would be in cusp profiles with slopes $\sim 1.4$, $\sim 1.5$, and $\sim 2$, respectively. In this case, the DF timescale would be even larger due to the small number of stars that move slower than the IMBH \citep{antonn2012}.}, and $\zeta(r)=M_{\rm cusp}(r)/\msmbh$ is the cusp mass at a given distance from the SMBH in units of the SMBH mass. At shorter distances, the inspiral is dominated by energy loss due to GW emission. In this case, an IMBH inspirals into the central SMBH on a timescale \citep{peters64},
\be
T_{\rm GW}=\frac{5}{256}\frac{r_h^4 c^5}{G^3 M_{\rm IMBH}^3 Q(1+Q)}\left(\frac{r}{r_h}\right)^{4}(1-e_{\rm IMBH}^2)^{7/2}\ ,
\label{eqn:tgw}
\ee
where $e_{\rm IMBH}$ is the IMBH orbital eccentricity.

In Figure~\ref{fig:timbhb}, we show the typical timescale (per IMBH) for the formation of IMBH--SBH at a given position within the cusp, for different IMBH and SMBH masses. Since the distribution of SBHs is cuspy, the smallest bremsstrahlung timescale occurs at the smallest galactocentric distance. Far from the SMBH ($r/r_h\gtrsim 0.1$), the typical timescale to form binaries can exceed Hubble time. On the other hand, $T_{\rm IMBHB}$ becomes of the order of $10^{-1}$--$10^{1}\,$Myr for smaller distances ($r/r_h\lesssim 10^{-3}$). The IMBH mass also affects the binary formation timescale. Since $\sigma_{\rm cap}\propto M_{\rm IMBH} \mathcal{R}_{\rm cap}$ (Eq.~\ref{eqn:crosect}) and $\mathcal{R}_{\rm cap}\propto M_{\rm IMBH}^{5/7}$ (Eq.~\ref{eqn:rss}), $T_{\rm IMBHB}\propto M_{\rm IMBH}^{-12/7}$. As a consequence, the bremsstrahlung timescale for a $5\times 10^3\msun$ IMBH is $\sim 10$ times longer than for a $2\times 10^4\msun$ IMBH. Moreover, $T_{\rm IMBHB}\propto M_{\rm SMBH}$, thus larger SMBH masses imply larger binary formation timescales.

We also report in Figure~\ref{fig:timbhb} the inspiral time due to dynamical friction and the GW merger timescale for the SMBH-IMBH system (assuming $e_{\rm IMBH}\sim 0$)\footnote{This assumption is motivated by the fact that the initial inspiral is mostly governed by dynamical friction which operates to circularize the IMBH orbit.}. An IMBH can capture SBHs whenever,
\be
T_{\rm IMBHB}< \min\ (T_{\rm DF},T_{\rm GW})\ .
\ee
For large orbital separations of the IMBH with respect to the central SMBH, dynamical friction is the main mechanism to lose energy, while GW energy loss operates on smaller distances. The typical distance $r_{\rm peak}$ at which $T_{\rm DF}=T_{\rm GW}$ is independent on the IMBH mass since both $T_{\rm DF}$ and $T_{\rm GW}$ are $\propto 1/M_{\rm IMBH}$. On the other hand, their slope depends on the SMBH mass and (for $T_{\rm DF}$) the cusp profile. The region where $T_{\rm IMBHB}$ is the smallest timescale corresponds to the available region of the phase space where IMBH--SBH binaries can efficiently form and eventually merge (shaded region in Fig.~\ref{fig:timbhb}). Actually, if an IMBH--SBH binary forms and $T_{\rm GW,IMBHB}> \max\ (T_{\rm DF},T_{\rm GW})$, the IMBH--SBH will inspiral into the SMBH producing a double LISA signal, essentially a superimposition of an IMRI (IMBH--SMBH inspiral) and an extreme-mass ratio inspiral (EMRI; SBH--SMBH inspiral).

We illustrate in Fig.~\ref{fig:timbhb}, the allowed parameter space for forming IMBH--SBH binaries, for $M_{\rm IMBH}=5\times 10^3\msun$ (top) and $M_{\rm IMBH}=2\times 10^4\msun$ (bottom). Smaller IMBHs have longer dynamical friction and GW timescales, thus they will inspiral onto the SMBH on longer timescales. However, since $T_{\rm IMBHB}\propto M_{\rm IMBH}^{-12/7}$, it is less probable for smaller IMBHs to capture SBHs. As a result, the size of the available region depends importantly on the IMBH mass. We find that $T_{\rm IMBHB}$ is never the smallest timescale for $M_{\rm IMBH}\lesssim 3\times 10^3\msun$. Heavier IMBHs can efficiently form binaries on short enough timescales. On the other hand, captured SBHs will typically have wide orbits, thus they are either orbitally unstable against SMBH perturbations (see next Section) or could merge with the IMBH on timescales longer than its inspiral time. We find that this is almost always the case for masses $\gtrsim 2\times 10^4\msun$.

In conclusion, gravitational bremsstrahlung is efficient in the IMBH mass range $3\times 10^3\msun\lesssim M_{\rm IMBH} \lesssim 2\times 10^4\msun$.

\section{Monte Carlo experiments}
\label{sect:montecarlo}

In this section we describe the Monte Carlo framework we developed to derive the distributions of the relevant parameters of binaries formed through the GW capture of an SBH by an IMBH, including their GW peak frequencies and strains.

As an illustrative example, in the following we consider the bremsstrahlung by a single IMBH. We consider the IMBH mass in the range $[5\times 10^3\msun$--$2\times 10^4\msun]$, and fix $m_2=10 \msun$. Our Monte Carlo routine is based on the following steps:
\begin{itemize}
    \item We draw randomly the galactocentric location $r$ in the interval $[r_{\min};r_{\max}]$ where the bremsstrahlung takes place, by accounting for that the event $\propto \min\ (T_{\rm DF},T_{\rm GW})/T_{\rm IMBHB}$. The minimum and maximum galactocentric distances are roots of the equations $T_{\rm IMBHB}(r_{\min})=T_{\rm GW}(r_{\min})$ and $T_{\rm IMBHB}(r_{\max})=T_{\rm DF}(r_{\max})$, respectively.
    \item At a given galactocentric distance, we sample the relative velocity in the range $[0$-$v_{\max} (r)]$, where,
    \be
    v_{\max}(r)=\left(\frac{8G\msmbh}{r}\right)^{1/2}\ .
    \ee
    \item We compute the maximum pericenter that results in a capture $\mathcal{R}_{\rm cap}(r)$ from Eq.~\ref{eqn:rss}.
    \item We sample the impact parameter in the range $b_{\min}$-$b_{\max}$ (Eq.~\ref{eqn:bmax}-\ref{eqn:bmin}), with a distribution $f(b)\propto b$. In doing this, we note that $b_{\min}<b_{\max}$ can be violated for high values of the relative velocity \citep{gondan2018}. This implies that these systems always suffer from a head-on collision.
    \item We require that the binary is tidally stable against perturbations by the SMBH,
    \begin{equation}
    r\gtrsim a\left(\frac{3\msmbh}{M_{\rm IMBH}}\right)^{1/3}\ ,
    \end{equation}
    where $a$ is the binary semi-major axis.
    \item We compute the semi-major axis ($a$), eccentricity ($e$), peak-frequency ($f_{\rm GW}$) and GW merger timescale ($T_{\rm GW}$) of the newly formed binary.
    \item The binary evolves according to \citep{peters64},
    \begin{equation}
    \frac{da}{dt}=-\frac{64}{5}\frac{G^3 M_{\rm IMBH} q (1+q)}{c^5 a^3 (1-e^2)^{7/2}}\left(1+\frac{73}{24}e^2+\frac{37}{96}e^4\right)\ ,
    \label{eqn:semgw}
    \end{equation}
    \begin{equation}
    \frac{de}{dt}=-\frac{304}{15}\frac{G^3 M_{\rm IMBH} q (1+q)}{c^5 a^4 (1-e^2)^{5/2}}\left(e+\frac{121}{304}e^3\right)\ ,
    \label{eqn:eccgw}
    \end{equation}
    and merges within $T_{\rm GW,IMBHB}$ (Eq.~\ref{eqn:tgwimbhb}).
\end{itemize}

\begin{figure} 
\centering
\includegraphics[scale=0.55]{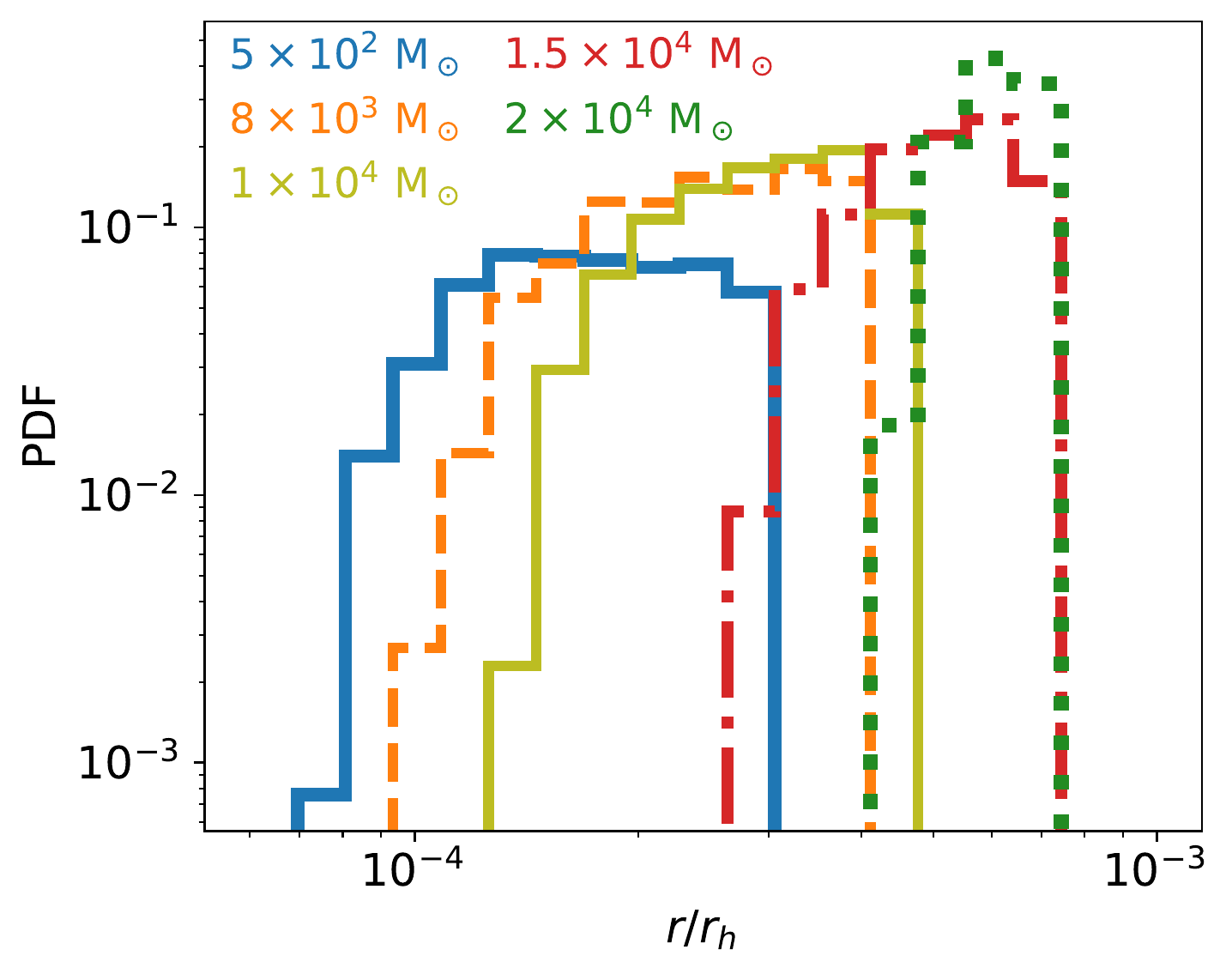}
\caption{Distribution of galactocentric locations of the IMBH--SBH binaries formed from the GW-capture of a $10\msun$ SBH, for different values of the IMBH mass. The mass of the SMBH is $\msmbh=4\times 10^6\msun$.}
\label{fig:rrh_imbh}
\end{figure}

In Fig.~\ref{fig:rrh_imbh}, we illustrate the distribution of galactocentric locations of the binaries formed from the GW-capture of a $10\msun$ SBH, for different values of the IMBH mass. The mass of the SMBH is $\msmbh=4\times 10^6\msun$. We find that the typical $r/r_h$ of the IMBH--SBH depends on the IMBH mass. The smaller the IMBH mass, the closer to the SMBH the IMBH--SBH binary forms and merge. This behavior is the result of two effects. First, the region of the parameter space where $T_{\rm IMBHB}< \min\ (T_{\rm DF},T_{\rm GW})$ is at smaller distances with respect to the SMBH for smaller IMBHs. Second, while heavier IMBHs can efficiently form binaries on short timescales in a wider area, captured SBHs will typically have wide orbits. As a result, they are either separated by the tidal field of the SMBH or inspiral into the IMBH on timescales longer than the IMBH inspiral time into the SMBH.

\begin{figure*} 
\centering
\includegraphics[scale=0.7]{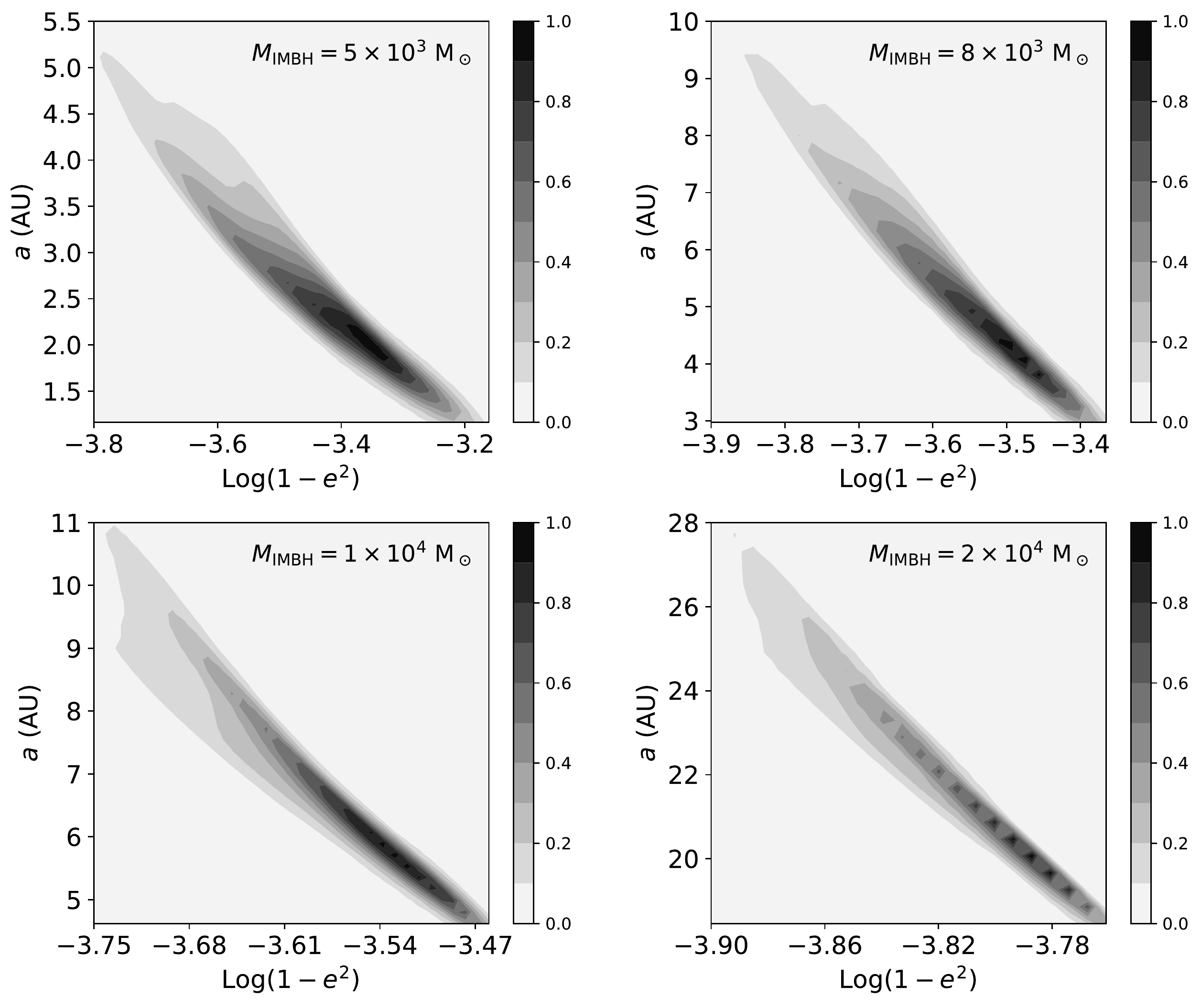}
\caption{Distribution of semi-major axis $a$ and $\epsilon=1-e^2$ of the IMBH--SBH of the binaries formed from the GW-capture of a $10\msun$ SBH, for different values of the IMBH mass. The mass of the SMBH is $\msmbh=4\times 10^6\msun$. Color bar: normalized probability density.}
\label{fig:ae_imbh}
\end{figure*}

We show the distribution of semi-major axis $a$ and $\epsilon=1-e^2$ of the IMBH--SBH, for $M_{\rm IMBH}=5\times 10^3\msun$, $8\times 10^3\msun$, $1\times 10^4\msun$, $2\times 10^4\msun$, in Fig.~\ref{fig:ae_imbh}. For any IMBH mass, smaller IMBH--SBH semi-major axes imply less eccentric orbits. We find that the distribution of semi-major axis is peaked at $\sim 2$ AU, $\sim 5$ AU, $\sim 7$ AU, $\sim 20$ AU for $M_{\rm IMBH}=5\times 10^3\msun$, $8\times 10^3\msun$, $1\times 10^4\msun$, $2\times 10^4\msun$, respectively, thus approximately $a\propto M_{\rm IMBH}^2$ (see also Eq.~\ref{eqn:semcap}). On the other hand, the typical value of $\epsilon$ is smaller for larger IMBH masses. We find that the distribution of $\epsilon$ values is peaked at $\sim 4\times 10^{-4}$, $\sim 3\times 10^{-4}$, $\sim 2\times 10^{-4}$, $\sim 1\times 10^{-4}$ for $M_{\rm IMBH}=5\times 10^3\msun$, $8\times 10^3\msun$, $1\times 10^4\msun$, $2\times 10^4\msun$, respectively. Thus, $\epsilon\propto M_{\rm IMBH}^{-1}$ (see also Eq.~\ref{eqn:ecccap}).

\begin{figure*} 
\centering
\includegraphics[scale=0.7]{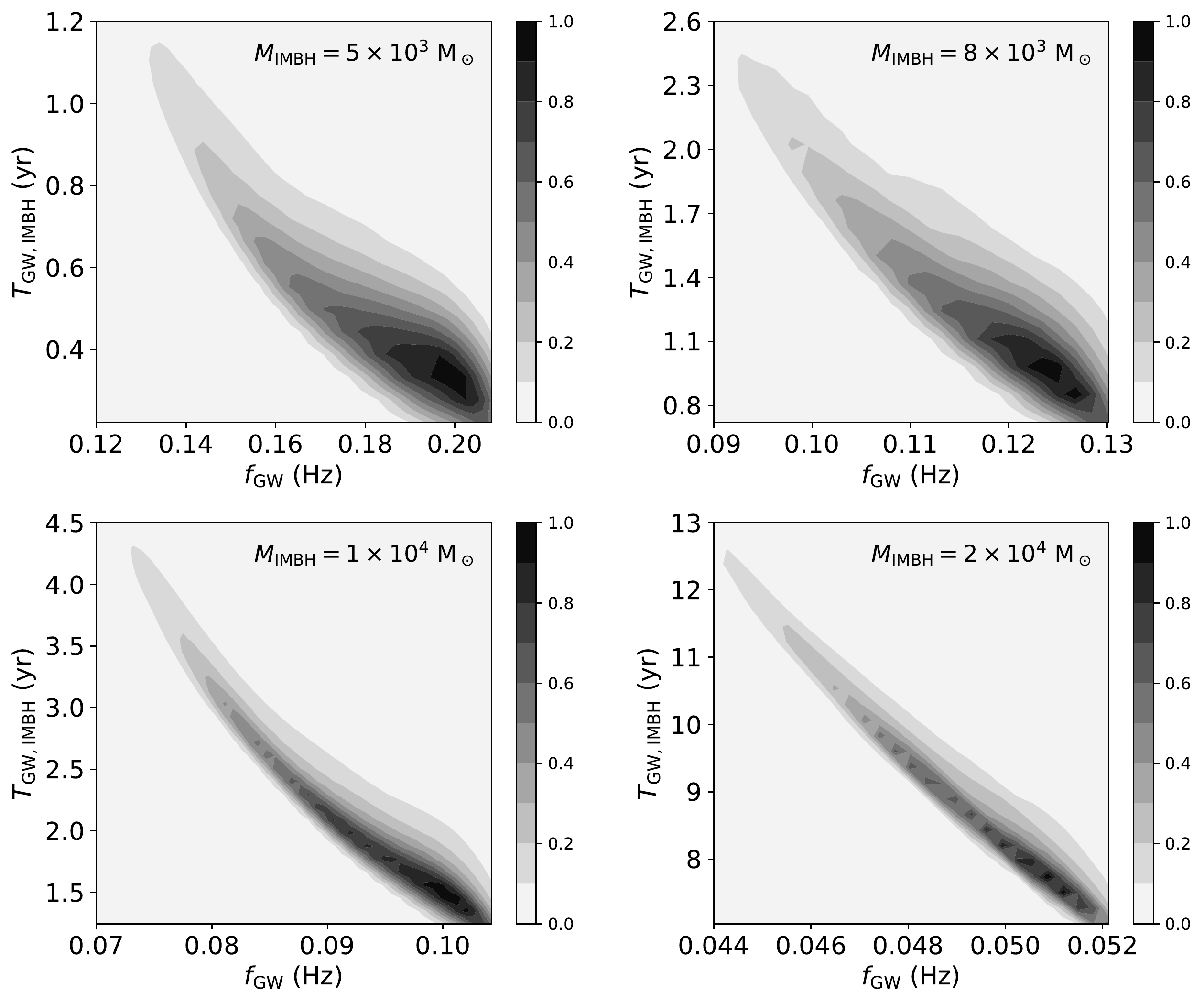}
\caption{Distribution of peak frequency and GW merger timescale of the binaries formed from the GW-capture of a $10\msun$ SBH, for different values of the IMBH mass. The mass of the SMBH is $\msmbh=4\times 10^6\msun$. Color bar: normalized probability density.}
\label{fig:tf_imbh}
\end{figure*}

In Figure~\ref{fig:tf_imbh}, we illustrate the distribution of peak frequency and $T_{\rm GW,IMBHB}$, for $M_{\rm IMBH}=5\times 10^3\msun$, $8\times 10^3\msun$, $1\times 10^4\msun$, $2\times 10^4\msun$. Different IMBH masses emit at different GW peak frequencies, the larger the IMBH mass the smaller the $f_{\rm GW}$. In particular, we find a peak at $f_{\rm GW}\sim 0.2\,$Hz, $0.1\,$Hz, $0.0.9\,$Hz, $0.05\,$Hz for $M_{\rm IMBH}=5\times 10^3\msun$, $8\times 10^3\msun$, $1\times 10^4\msun$, $2\times 10^4\msun$, respectively. The distribution of GW merger timescales is peaked at $\sim 0.3\,$yr, $\sim 1\,$yr, $\sim 2\,$yr, $\sim 10\,$yr for $M_{\rm IMBH}=5\times 10^3\msun$, $8\times 10^3\msun$, $1\times 10^4\msun$, $2\times 10^4\msun$, never long enough for external secular perturbations to matter. Note that this can be explained by considering,
\be
T_{\rm GW,IMBHB}\propto \frac{a^4 \epsilon^{7/2}}{M_{\rm IMBH}^2}\propto M_{\rm IMBH}^{5/2}\ .
\ee
While for SBH GW-captures $T_{\rm GW}$ is of the order of seconds \citep{oleary2009}, thus resulting in a rapid GW signal, the merger timescale is of the order of minutes up to years for the IMBH regime.

\section{Multiband gravitational wave observations}
\label{sect:multiband}

We are now in the position to describe the typical GW signal expected from binaries merging as a result of the IMBH bremsstrahlung process.

\begin{figure*} 
\centering
\includegraphics[scale=0.7]{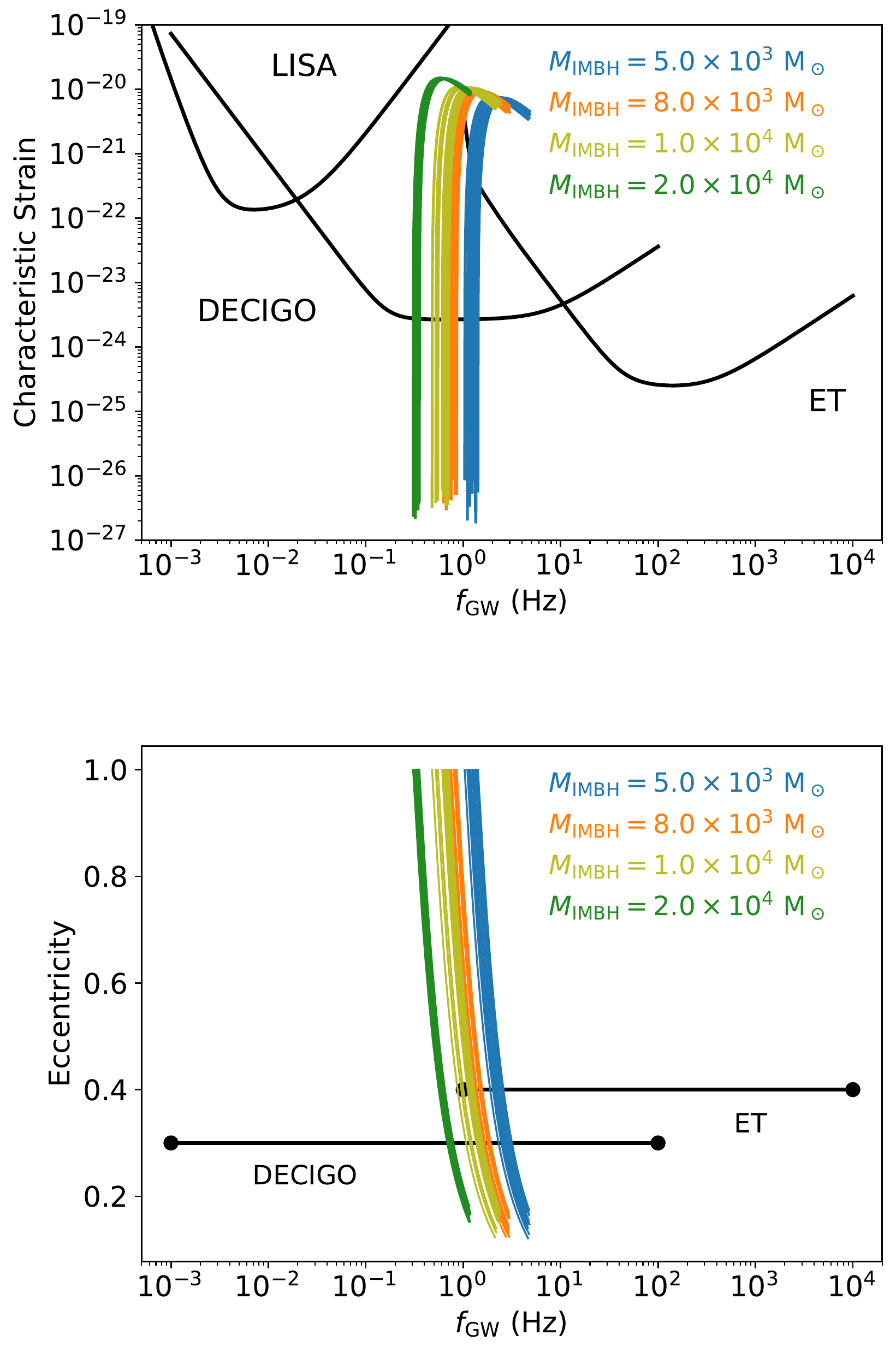}
\caption{Evolution of the characteristic strain (top) and eccentricity (bottom) at frequency of peak emission for different values of the IMBH mass. We assume a distance of $250$ Mpc from Earth. Black curves represent the ET \citep{ETSensitivity}, DECIGO \citep{yagi2011}, and LISA \citep{robson2019} sensitivity curves.}
\label{fig:strain}
\end{figure*}

For an eccentric binary, the characteristic strain at the $n$-th armonic can be written as \citep{barack2004},
\be
h_{\rm c,n}^2=\frac{1}{(\pi D)^2}\left(\frac{2G}{c^3}\frac{\dot{E}_{\rm n}}{\dot{f}_{\rm n}}\right)\ ,
\label{eqn:nstrain}
\ee
where,
\be
f_{\rm n} = nf_{\rm orb}\ ,
\ee
where $f_{\rm orb}$ is the orbital frequency. This is related to the observed (detector frame) frequency $f_{\rm n,z}$ by $f_{\rm n}=f_{\rm n,z}(1+z)$. In Eq.~\ref{eqn:nstrain}, $\dot{E}_{\rm n}$ is the time derivative of the energy radiated by GWs at the frequency $f_{\rm n}$ \citep{pet1963},
\be
\dot{E}_{\rm n}=\frac{32}{5}\frac{G^{7/3}}{c^5}(2\pi f_{\rm orb})^{10/3} g(n,e)\ ,
\label{eqn:endot}
\ee
where $M_\mathrm{c}$ is the rest-frame chirp mass,
\be
M_\mathrm{c}=M_\mathrm{c,z}(1+z)=M_{\rm IMBH} \frac{q^{3/5}}{(1+q)^{1/5}}(1+z)
\ee
and $g(n,e)$ is a combination of Bessel functions of the first kind \citep[see Eq. 20 in][]{pet1963}. Using $\dot{f}_{\rm n} = n\dot{f}_{\rm orb}$ and the semi-major axis Peters' equation \citep{peters64},
\be
\frac{da}{dt}=-\frac{64}{5}\frac{G^3 M_{\rm IMBH}^3 q (1+q)}{c^5 a^3} F(e)\ ,
\ee
the derivative of the $n$-th harmonic can be written as,
\be
\dot{f}_{\rm n}=n\frac{96}{10\pi}\frac{(GM_{\rm c})^{5/3}}{c^5}(2\pi f_{\rm orb})^{11/3}F(e)\ ,
\label{eqn:fndot}
\ee
where,
\be
F(e)=\frac{1+(73/24)e^2+(37/96)e^4}{(1-e^2)^{7/2}}\ .
\ee
Combining Eq.~\ref{eqn:endot}-\ref{eqn:fndot} and Eq.~\ref{eqn:nstrain}, the characteristic strain at the $n$-th harmonic can be rewritten as,
\be
h_{\rm c,n}^2=\frac{2}{3\pi^{4/3}}\left(\frac{2}{n}\right)^{2/3}\frac{G^{5/3} M_{\rm c}^{5/3}}{f_{\rm n,z}^{1/3} (1+z)^2 c^3 D^2}\frac{g(n,e)}{F(e)}\ .
\ee

In the top panel of Figure~\ref{fig:strain}, we show the evolution of the characteristic strain at frequency of peak emission for different values of the IMBH mass, assuming a distance of $250$ Mpc from Earth. We also show for comparison the sensitivity curves for ET \citep{ETSensitivity}, DECIGO \citep{yagi2011}, and LISA \citep{robson2019}. Since we find that the typical peak frequency is $\sim 0.2\,$Hz, $0.1\,$Hz, $0.09\,$Hz, $0.05\,$Hz for $M_{\rm IMBH}=5\times 10^3\msun$, $8\times 10^3\msun$, $1\times 10^4\msun$, $2\times 10^4\msun$, respectively, merging IMBH of different masses will appear in the sensitivity frequency band of different instruments. Low-mass IMBHs typically will first appear in the DECIGO band and than will be observed also by ET, as they inspiral towards the merger. More massive IMBHs ($\gtrsim 10^4\msun$) will appear only in DECIGO. 

We illustrate the eccentricity evolution at frequency of peak emission in the bottom panel of Figure~\ref{fig:strain}. We also plot the frequency range at which ET and DECIGO can measure IMBH--SBH inspirals. Some of the mergers still retain a non-negligible eccentricity when they enter the detector frequency band, which will be typically $\gtrsim 0.1$. Measuring the retained eccentricity, using eccentric waveform templates, in particular when the inspiral can be detected by different instruments, would shed light on the formation scenario and reveal binaries formed as a result of gravitational bremsstrahlung in galactic nuclei. This holds true for IMBH binaries formed in the center of globular clusters \citep{mandel2008}.

\section{Rates}
\label{sect:rates}

Next, we provide a simple estimate of the expected rate of GW events from IMBH bremsstrahlung.

The predominant population of IMBH--SBH binaries is formed at a typical distance ($r_{\rm peak}$) from the SMBH where the GW or dynamical friction timescale is the longest and $T_{\rm IMBHB}$ is the shortest. Therefore, typical number $N_{\rm IMBHB}$ of formed IMBH--SBH through captures is,
\be
N_{\rm IMBHB}\sim \max_{r_{\min}<r<r_{\max}} \frac {\min\ (T_{\rm DF},T_{\rm GW})}{T_{\rm IMBHB}}\sim \frac {T_{\rm GW}(r_{\rm peak})}{T_{\rm IMBHB}(r_{\rm peak})}\ ,
\label{eqn:timbhb}
\ee
where $r_{\rm peak}$ is the distance at which $T_{\rm DF}(r_{\rm peak})=T_{\rm GW}(r_{\rm peak})$. From Fig.~\ref{fig:timbhb}, $N_{\rm IMBHB}\sim 3$--$5$ over a timescale $T_{\rm GW}(r_{\rm peak})\sim 10$-$15\,$yr for a $5\times 10^3\msun$ IMBH in a Milky Way-like nucleus. For a $2\times 10^4\msun$ IMBH, $N_{\rm IMBHB}\sim 6$--$10$. Therefore, the rate of IMBH--SBH mergers can be as high as $\sim$ few yr$^{-1}$ during the SMBH-IMBH inspiral.

The above numbers have been derived per IMBH hosted in a given nucleus at a given time. Of course, IMBHs can inspiral onto the SMBH due to the combined effect of dynamical friction (Eq.~\ref{eqn:tdf}) and GW emission (Eq.~\ref{eqn:tgw}) producing a GW signal observable by LISA up to large redshifts \citep{agu18}. Eventually the formation rate $\Gamma_{\rm form}$ of IMBHs (through the processes described in Section~\ref{sect:formation}) would be large enough to replenish the innermost galactic regions with newly-formed IMBHs. The Milky Way galactic center may host several IMBHs in its nuclear star cluster, whose dynamical effects and/or nHz-frequency GW may be detected in the future. These considerations, along with constrains from the orbital stability of S-stars \citep{gual2009,naoz2019} and proper motion measurements of Sgr A$^*$ \citep{hansen2003,reid2004}, have been used to constrain the possible IMBH companion to the SMBH in our Galactic Center within the central parsec. Interestingly, the relevant range of IMBH masses for GW captures in a Milky Way-like nucleus overlaps with the allowed parameter space for a secondary massive black hole in our Galactic Center. Therefore, monitoring Sgr A$^*$ with ET and DECIGO could place tighter constraints on the possible secondary IMBH, companion to Sgr A$^*$.

If the formation rate is larger than the merging rate, $\Gamma_{\rm form}\gtrsim 1/\min\ (T_{\rm DF},T_{\rm GW})$, more than one IMBH can accumulate in a given galactic nucleus. In the calculation above, we have considered the bremsstrahlung process from a single IMBH. If more IMBHs are present within the SMBH sphere of influence, we could in principle apply the same procedure outlined above for all of them. In this case, it would be important to also compute the typical timescale for $2$ IMBHs interacting, which could affect the rate estimate to some extent. Assuming a distribution of IMBHs as,
\be
n_{\rm IMBH}=\frac{N_{\rm IMBH}}{A r_h^3}\left(\frac{r}{r_h}\right)^{-\alpha_{\rm IMBH}}\ ,
\ee
where $N_{\rm IMBH}$ is the number of IMBHs, the IMBH-IMBH interaction timescale would be,
\be
T_{\rm 2IMBH}=\frac{1}{N_{\rm IMBH} n_{\rm IMBH} \sigma_{\rm cap} v}\ .
\ee
Interestingly, this mechanism can create IMBH binaries, which can later merge and can possibly be kicked out by GW recoil kicks \citep{oll12}. We leave more detailed calculations of these effects to a future work.

\section{Discussion and conclusions}
\label{sect:conc}

IMBHs are one of the unsolved puzzles of modern astronomy with no conclusive evidence for their existence. They attract much interest owing to their important role in a wide range of phenomena, including the origin of SMBH seeds and galaxy evolution, tidal disruption events, dwarf galaxy feedback, and hypervelocity stars. 

In this paper, we have described and outlined for the first time the characteristics of the GW sources produced through IMBH--SBH captures in galactic nuclei. We have shown that the typical semi-major axis, eccentricity, peak GW frequency and merger timescales of the IMBH--SBH binaries formed as a result of this process depend mainly on the IMBH mass. In particular, we have found that the typical peak frequency is $\sim 0.2\,$Hz, $0.1\,$Hz, $0.09\,$Hz, $0.05\,$Hz for $M_{\rm IMBH}=5\times 10^3\msun$, $8\times 10^3\msun$, $1\times 10^4\msun$, $2\times 10^4\msun$, respectively. As such, low-mass IMBHs will typically appear in both DECIGO and ET bands, while more massive IMBHs only in DECIGO as they merge. Interestingly, while the merger timescales is of the order of seconds for SBH GW-captures \citep{oleary2009}, thus resulting in a rapid GW inspiral signal, it is of order months to years for the IMBH regime. Some of the mergers will appear eccentric in the detector frequency band.

As in the process described here for SBHs, IMBHs can also capture neutron stars, main-sequence stars, and white dwarfs. For white dwarfs, through the strong tidal interaction, this mechanism could trigger a thermonuclear explosion. The consumption of a white dwarf would be extremely interesting, since these events are luminous only for IMBH with masses $\lesssim 10^5\msun$ \citep{rossw2008,rossw2009,macl2016}.

Finally, we note that a similar calculation can be done for IMBH gravitational bremsstrahlung outside the SMBH influence radius and in galactic nuclei that do not host SMBHs. The latter might be the case of galaxies with mass $\lesssim 10^9 \msun$ \citep{ferr2006,capu2017}. The same exact process would be relevant in globular clusters hosting an IMBH in their center \citep{mandel2008}.

By measuring the mass, spin, and redshift distributions for IMBH--SBH mergers, next-generation GW observations may help to improve our understanding of galaxy formation and galactic nuclei.

\section*{Acknowledgements}

We thank the referee for a constructive report. GF acknowledges support from a CIERA Fellowship at Northwestern University. This work was supported in part by Harvard's Black Hole Initiative, which is funded by grants from JFT and GBMF. KK and FAR acknowledge support from NSF Grant AST-1716762.

\bibliographystyle{yahapj}
\bibliography{refs}

\end{document}